\documentclass[twoside,twocolumn,9pt]{article}
\usepackage{extsizes}
\usepackage[super,sort&compress,comma]{natbib} 
\usepackage[left=1.5cm, right=1.5cm, top=1.785cm, bottom=2.0cm]{geometry}
\usepackage{balance}

\usepackage{sectsty}
\allsectionsfont{\scshape}

\usepackage{graphicx} 
\usepackage{lastpage}
\usepackage{float}
\usepackage{fancyhdr}
\usepackage{fnpos}
\usepackage[english]{babel}
\usepackage{array}
\usepackage[T1]{fontenc}
\usepackage[usenames,dvipsnames]{xcolor}
\usepackage{setspace}
\usepackage{hyperref}
\usepackage{subfigure}

\usepackage{upgreek}
\usepackage{xspace}
\usepackage{placeins}
\usepackage{amsmath}
\usepackage{amssymb}

\usepackage{mciteplus}
\usepackage{adjustbox}
\usepackage{epstopdf}
\usepackage{stfloats}

\begin{document}

\twocolumn[
  \begin{@twocolumnfalse}
\vspace{3cm}

\begin{center}

    \noindent\huge{\textbf{\textsc{Clog mitigation in a microfluidic array via pulsatile flows}}} \\
    \vspace{1cm}

    \noindent\large{Brian Dincau,\textit{$^{a}$} Connor Tang,\textit{$^{a}$} Emilie Dressaire,\textit{$^{a}$} Alban Sauret \textit{$^{a}$}}$^{\ast}$ \\

    \vspace{5mm}
    \noindent\large{\today} \\

    \vspace{1cm}
    \textbf{\textsc{Abstract}}
    \vspace{2mm}

\end{center}

\noindent\normalsize{
    Clogging is a common obstacle encountered during the transport of suspensions and represents a significant energy and material cost across applications, including water purification, irrigation, biopharmaceutical processing, and aquifer recharge. Pulsatile pressure-driven flows can help mitigate clogging when compared to steady flows. Here, we study experimentally the influence of the amplitude of pulsation $0.25\,P_0 \leq \delta P \leq 1.25\,P_0$, where $P_0$ is the mean pressure, and of the frequency of pulsation $10^{-3}\,{\rm Hz} \leq f \leq 10^{-1}\,{\rm Hz}$ on clog mitigation in a microfluidic array of parallel channels using a dilute suspension of colloidal particles. The array geometry is representative of a classical filter, with parallel pores that clog over time, yielding a filter cake that continues to grow and can interact with other pores. We combine flow rate measurements with direct visualizations at the pore scale to correlate the observed clogging dynamics with the changes in flow rate. We observe that all pulsatile amplitudes at $0.1$ Hz yield increased throughput compared to steady flows. The rearrangement of particles when subject to a dynamic shear environment can delay the clogging of a pore or even remove an existing clog. However, this benefit is drastically reduced at $10^{-2}$ Hz and disappears at $10^{-3}$ Hz as the pulsatile timescale becomes too large compared to the timescale associated with the clogging and the growth of the filter cakes in this system. The present study demonstrates that pulsatile flows are a promising method to delay clogging at both the pore and system scale.} \\

 \end{@twocolumnfalse} \vspace{0.6cm}

  ]

\makeatletter
{\renewcommand*{\@makefnmark}{}
\footnotetext{\textit{$^{a}$~Department of Mechanical Engineering, University of California, Santa Barbara, California 93106, USA}}
\footnotetext{\textit{$^{*}$ asauret@ucsb.edu}}
\makeatother


\section{Introduction}

Clogging is a multiscale phenomenon present in nearly all confined suspension flows. Depending on the suspension properties and constriction geometry, suspended particles can deposit at a constriction, forming an aggregate that grows over time. The particles can also form a bridge, or even directly sieve the constrictions.\cite{Dressaire2017a} Clogging and the resulting growth of the filter cake reduce the permeability of the constriction, substantially reducing the flow rate at a given pressure.\cite{Sauret2018} Clogging can be dangerous, manifesting in life-threatening conditions like thrombosis,\cite{Mohanty1994VascularMalaria,Ye2020TheAggregation} or natural disasters like flooding.\cite{Kia2017CloggingReview} Clogging is also a significant issue in water management, affecting critical applications like groundwater recharge, \cite{katznelson1989clogging, Jeong2018AApplications} irrigation, \cite{Bounoua2016UnderstandingApproaches, Ait-Mouheb2019ImpactDripper} and desalination.\cite{Kashyap2017ALocalization} Besides, the clogging of colloidal microplastics is becoming increasingly important, as they are heavily utilized in biopharmaceutical manufacturing \cite{Giri2013ProspectsDelivery, c2015delivery, Dincau2017} and are an increasing environmental threat due to the erosion of plastic pollution.\cite{AlHarraq2021MicroplasticsScience} Furthermore, systems which have fully clogged and thus have a flow rate close to zero are often prohibitively challenging to unclog, and require a significant energy or material cost to restore. Changing a filter or backflushing a pipeline both require temporary inoperation of the system which they are a part of, as well as labor and materials to remedy the clog.

Clogging mechanisms are categorized according to the scenarios from which they arise. The nature of clogging mainly depends on the particle size $d$, the constriction size $L$, the volume fraction of the suspensions, and can also depends on the geometry of the channel.\cite{Dressaire2017a,bacchin2014clogging} Sieving describes the scenario in which $L<d$, thus particles cannot pass through the constriction. This is often the goal in filtration where particles above a certain size threshold determined by the filter pore size $L$ are removed. Sieving is typically the final event occurring in most clogging scenarios. \cite{Delouche2020DynamicsAggregates,Delouche2021TheScale} Bridging describes when several particles simultaneously arrive at a constriction and form a bridge due to geometric confinement. Bridging can rapidly block a constriction, but typically only occurs for concentrated enough suspensions when $L/d \lesssim 3 $ for non-adhesive and rigid spherical particles.\cite{Marin2018,Souzy2020TransitionSuspensions,hsu2021roughness,bielinski2021squeezing} Aggregation describes the successive deposition of suspended particles at a constriction, driven by particle-wall interactions.\cite{Wyss2006,Dersoir2015,dersoir2017clogging,Dersoir2019} Aggregation allows even large constrictions with $L \gg d$ to become clogged over time as small particles deposit and shrink the constriction, increasing the probability of other clogging mechanisms.\cite{Bizmark2020MultiscaleMedia} Overall, the clogging of practical systems often involves a combination of the three mechanisms.

\begin{table*}[bp]
\begin{adjustbox}{width=1\textwidth}
\begin{tabular}{lccccc}
\multicolumn{1}{c}{Reference} & Frequency (Hz) & Amplitude             & \begin{tabular}[c]{@{}c@{}}Main clogging\\ mechanism\end{tabular} & Geometry                         & Application               \\ \hline
Jackson \textit{et al.} (1987)\cite{Jackson1987}         & $10^{-4}$      & $800\%\,P_0$          & –                                                                 & Single pores                     & Irrigation                \\
Bichara \textit{et al.} (1988)   \cite{bichara1988redevelopment}      & –              & $200\%\,P_0$          & Mixed                                                             & Packed sediment bed              & Aquifer redevelopment     \\
McFaul \textit{et al.} (2012)   \cite{McFaul2012}       & $0.25$         & $150\%\,P_0$          & Sieving                                                           & Microarray                       & Cell separation           \\
Cheng \textit{et al.} (2016)    \cite{Cheng2016}       & –              & $>100\%\,P_0$         & –                                                                 & Microporous membrane             & Whole blood separation    \\
Yoon \textit{et al.} (2016)    \cite{Yoon2016a}        & $70-230$       & –                     & Sieving                                                           & Microarray                       & Cancer detection          \\
Zhang \textit{et al.} (2017)   \cite{Zhang2017}        & 0.04           & $70\%\,P_0$           & Aggregation                                                       & Microarray and labyrinth channel & Irrigation                \\
Lee \textit{et al.} (2018)        \cite{Lee2018a}     & 0.125          & –                     & –                                                                 & Microarray                       & Blood plasma filtration   \\
Cheng \textit{et al.} (2018)     \cite{Cheng2018a}      & 1              & Oscillatory ($P_0=0$) & Sieving                                                           & Microarray                       & Cell capture \& staining \\
Mehendale \textit{et al.} (2018)  \cite{Mehendale2018}     & –              & –                     & –                                                                 & Micropillar array                & Whole blood enrichment    \\
Kastl \textit{et al.} (2021)   \cite{kastl2021impact}        & $5-10$         & $35-80\%\,P_0$        & Aggregation                                                       & Microporous membrane             & Forward osmosis           \\ \hline
\end{tabular}
 \end{adjustbox}
\caption{\label{tab:Table 1}Summary of some studies that utilized pulsatile flows for clog mitigation and relevant parameters, main clogging mechanism, geometry of the channels, and applications considered.}
\end{table*}

Currently, clog prevention is primarily handled in two ways.\cite{aljuboury2017treatment,hoslett2018surface,hube2020direct,sikka2021critical} The first approach is to rely on an upstream filtration, wherein particles are captured at a known, accessible location upstream of susceptible constrictions. A second common approach is to rely on chemical addition, wherein acids, bases, or biocides like chlorine are used to eliminate suspended particles or aggregates.\cite{trooien1998filtration,Goyal2015} However, both of these techniques have clear disadvantages. Upstream filtration prevents clogging later down the line, but filters themselves may clog relatively quickly, and filter replacement has a cost. Chemical addition is limited by the material properties of the system, and comes with the long-term risks of altered biodiversity \cite{Bertelli2018ReducedBiofilms} and chemical invasion \cite{EINAV2003141} into nearby ecosystems. In fact, chemical addition can preferentially select for opportunistic microorganisms while eliminating their competition, causing systems to clog even faster.\cite{Wang2019} For these reasons, there is a strong motivation to develop active or passive hydrodynamic techniques to prevent or delay clogging.

In particular, while bridging and sieving can sometimes be mitigated with proper filtration and dilution, aggregation remains an intricate clogging mechanism to prevent in many applications. For instance, bacteria and algae are well-known clogging sources in irrigation, because the individual cells are too small to filter practically and can attach at constrictions via aggregation. Attached bioparticles can feed off of nutrients in the flow, allowing them to grow from micro-scale clusters to 3D milli-scale communities that clog the constriction.\cite{Goyal2015, Drescher2013BiofilmSystems} Besides, aggregation is present in any suspension flow where particles are small enough for Van der Walls forces to play a role. \cite{Lebovka2014AggregationParticles} As a result, various past studies have investigated particle deposition and aggregation at the micro-channel scale \cite{Wyss2006} and the pore-scale,\cite{Dersoir2015} some using Derjaguin–Landau–Verwey–Overbeek (DLVO) theory to explain their findings. These studies utilize steady flows with a constant driving pressure, which is typical of many practical flows. Recently, Trofa \textit{et al.} numerically investigated clogging under both constant pressure and constant flow rate for adhesive particles.\cite{Trofa2021NumericalContraction} They demonstrated the potential to predict numerically fouling and clogging for steady flows, given sufficient input physical parameters. 

Schwarze \textit{et al.} investigated particle attachment and detachment under an accelerating shear flow. They observed a decrease in net particle deposition with increasing shear, as well as a rate-dependent hysteresis.\cite{Schwarze2019AttachmentFlow} In their experiments, particles that deposit and remain attached at a lower shear can detach and become re-suspended at higher shear. In fact, hydrodynamic techniques like backflushing and high-pressure flushing have historically been used to periodically manage clogging, as reversing the shear direction or increasing its magnitude can shed particles from partially clogged constrictions.\cite{zsirai2012efficacy,lohaus2020microscopic} These observations suggest that this phenomenon can be leveraged quantitatively by continually varying the shear environment via pulsatile flows.

A pulsatile pressure-driven flow can be expressed using the instantaneous driving pressure $P(t) = P_0 + \delta P\,\sin(2\,\pi\,f\,t)$, where $P_0$ is the mean pressure, $\delta P$ is the pulsatile amplitude, $t$ is time, and $f$ is the pulsatile frequency.\cite{dincau2019pulsatile} Pulsatile flows have been proposed as a strategy to help mitigate clogging across applications spanning irrigation, water purification, blood filtration, and cancer detection.\cite{Jackson1987,bichara1988redevelopment,McFaul2012,Cheng2016,Yoon2016a,Zhang2017,Lee2018a,Cheng2018a,Mehendale2018,kastl2021impact} Table \ref{tab:Table 1} summarizes several studies that utilize pulsatile flows to mitigate clogging in a system. The frequency $f$, amplitude $\delta P$, primary clogging mechanism, constriction geometry, and application are noted if present in the original study. Most of these studies feature similar constriction geometries, \textit{i.e.}, a parallel configuration of microchannels or micropores. There is a drastic range for frequencies and amplitudes across these experiments, despite many of them operating at similar scales. Although the studies mentioned in Table 1 demonstrate clog mitigation via pulsation, none are systematic with respect to pulsation and they often fail to report critical pulsatile parameters. Furthermore, most studies focus on either the pore-scale or system-scale thus lacking the information to bridge the microscale dynamics to the mascroscopic flow rate. In particular, no mechanisms to explain why pulsatile flows may be efficient at delaying clogging were clearly visualized. Thus there remains a significant knowledge gap in optimizing pulsatile parameters and pumping schemes to balance clog mitigation with energy input. Besides, to better understand the mechanisms at play during the clogging of microchannels in pulsatile flow, it is important to be able to visualize the dynamics both at the macroscopic scale but also at the pore scale.

In this study, we investigate experimentally a range of pulsatile amplitudes $\delta P$ and frequencies $f$ as they relate to clog mitigation in a microfluidic array. We first present the experimental methods in section \ref{sec:Experiments}. The array, shown in figure \ref{fgr:Figure_1}, shares common features with a conventional filter in geometry, consisting of identical parallel pores. We enforce the instantaneous driving pressure $P(t)$ in our experiments while directly measuring the total flow rate $Q(t)$ through the array.  Section \ref{sec:Results} then reports the experimental results for both clogging under steady conditions and for pulsatile flow when varying the amplitude and the frequency of pulsations. We combine the flow rate measurements with direct visualization to correlate clogging at the individual pore scale with measurement of the flow rate at the device scale. Finally, section \ref{sec:discussion} considers in more detail the change in the probability of clogging with pulsatile flows. In particular, we show that an exponential distribution, corresponding to a Poisson process, captures the clogging dynamics in both steady and pulsatile conditions, but the time or volume intervals between two clogging events increases for a pulsatile flow. We also highlight some clogging and unclogging scenarios that explain our observations and shed light on the mechanisms behind pulsatile clog mitigation.


\section{Experimental Methods} \label{sec:Experiments}

\begin{figure}
\centering
\includegraphics[width=0.5\textwidth]{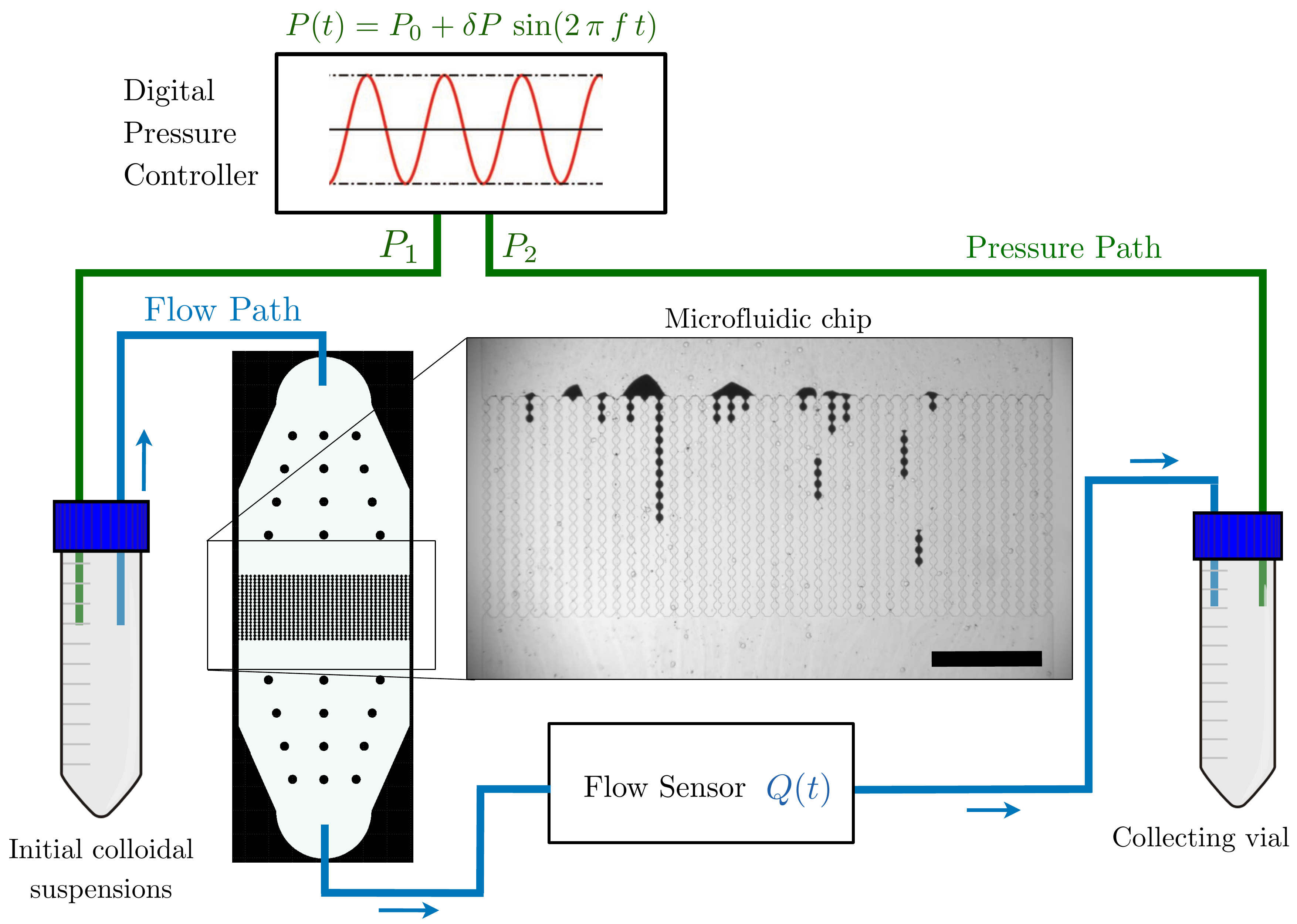}
  \caption{Schematic of the experimental setup showing the pressure driven system, the microfluidic device and the flow sensor. The square area indicates the region of observation. Scale bar is $1$ mm.}
  \label{fgr:Figure_1}
\end{figure}

\subsection{Microfluidic chip}

The microfluidic array, shown in figure \ref{fgr:Figure_1} is fabricated out of polydimethylsiloxane (PDMS) Sylgard 184 through soft lithography and plasma-bonding to a PDMS-coated slide. This method ensures that all internal surfaces are made of the same material. Plasma-bonding leaves the surfaces temporarily hydrophilic, and they slowly recover their hydrophobicity over time. Uncontrolled hydrophobicity could cause different clogging rates between experiments.\cite{Cejas2017ParticleStrength} To address this issue, all devices are thermally annealed at $120\; ^{\rm o}$C for at least three hours prior to experiments to ensure consistent hydrophobicity.\cite{Dersoir2015} The array consists of 40 microchannels, each with minimum constrictions of $10\, \mu{\rm m}$ and largest width of $50\, \mu{\rm m}$ and a channel depth of 13 to 15 $\mu$m.\cite{Wyss2006,sauret2014clogging} The parallel channels connect two large reservoirs, which serve as the inlet and outlet for the suspension flow. Each reservoir has 12 large pillars which serve as anchors to minimize pressure-induced deformation. Deformation is undesirable in these experiments, as it makes the pressure-flow rate relationship nonlinear, and could also cause particles attached in the reservoir to shift and migrate down to the array. Since the nonlinearity between $P$ and $Q$ would influence clogging in an uncontrolled manner, we verified experimentally that we work in the linear regime, even for the largest pressure difference used during the pulsatile experiments.

\subsection{Suspension properties and preparation}

The suspension is comprised of pure water (Millipore Milli-Q), 100 mM NaCl (Sigma-Aldrich), and $2 \,\mu{\rm m}$ carboxylate-modified latex beads (Invitrogen) at a volume fraction of $\phi = 0.3\%$. At this volume fraction, the suspension viscosity remains unchanged from water,\cite{mewis2012colloidal} while also minimizing the probability of bridging events.\cite{Marin2018} The beads have a density $\rho=1.055$ g cm$^{-3}$ at $20 \; ^{o} $C, thus sedimentation is negligible. The particles are stable up to 1 M univalent salt, and even higher based on our testings. Without the addition of salt, we encountered variation in clogging times that we believe may be due to streaming electrification, which can occur when ultra-pure water flows past insulating surfaces. \cite{Matsui1993ElectrificationSheet,Yatsuzuka1996ElectrificationWater} The addition of salt prevents charge accumulation and improves the repeatability of experiments. Besides, the addition of salt is a common aspect of numerous clogging studies, as the increased ion availability decreases the Debye's length and pushes the aggregation process into the Van der Waals Regime.\cite{mustin2016single,Cejas2017ParticleStrength}

Each suspension is filtered through an 8 $\mu{\rm m}$ track-etched membrane filter (Whatman Nuclepore) directly before experiments. This eliminates particles and aggregates that would immediately clog the array through sieving since the width of the constriction is $W=10\,\mu{\rm m}$. We measured the particle size distribution and obtained a resulting diameter $d=2.4 \pm 0.56 \,\mu{\rm m}$  (see Supplemental Materials), smaller than the constriction width $W$ with an aspect ratio $W/d \simeq 4.2$ thus, most particles cannot lead to clogging by sieving or bridging.\cite{Dressaire2017a} However, larger anisotropic aggregates in the tail of the distribution could sieve the constriction after a monolayer of particles deposit and shrink the local width of the constriction. During the experiment, we avoid stirring the suspension to minimize shear stress prior to reaching the array. The velocity field associated with stirring increases the probability of particle interaction, which could cause particles to aggregate prior to reaching the array. 

\subsection{Experiment overview}

A diagram of the entire experimental setup is depicted in figure \ref{fgr:Figure_1}. For each experiment, inlet and outlet pressure are enforced by a digital pressure controller (Elveflow OB1 MK3+) with $P_0=150\,{\rm mbar}$. The average pressure $P_0$ is kept constant in all the experiments presented in this paper, and the amplitude and frequency of pulsation are varied in the range $37.5\,{\rm mbar} \leq \delta P \leq 187.5\,{\rm mbar}$ and $10^{-3}\,{\rm Hz} \leq f \leq 10^{-1}\,{\rm Hz}$, respectively. These pulsations are directly imposed through a commercial pressure controller, which relies on piezo regulation. We tested the system frequency limitation, \textit{i.e.}, the maximum frequency at which the flow rate still follows the time-varying pressure to select the upper frequency limit, $f = 0.1\,{\rm Hz}$, for our system (see Supplemental Materials). The flow rate is continually measured with an in-series Coriolis flow sensor (Bronkhorst) downstream of the array. The flow sensor smooths the measurement over intervals smaller than 1 s. However, since the raw data exhibits some fluctuations inherent to the accuracy of the flow senser, we further smooth the data using a custom-made MATLAB routine based on an moving average with a window of typically $100\,{\rm s}$ for $f=0.01\,{\rm Hz}$ and of $50\,{\rm s}$ for every other cases. This time window is smaller than the clogging time and thus does not influence our measurements of the clogging frequency and the long-term evolution of the flow rate. However, we shall see later that for the smallest frequency used in this study, $f=0.001\,{\rm Hz}$, the smoothing interval is such that we still observe the oscillations in the flow rate. The array is transparent and positioned under an inverted microscope (Nikon Eclipse), allowing us to record a video at two frames per second with the attached USB camera (IDS Imaging) for the entire duration of each experiment, typically around 10 hours.

For each experiment, we begin by pre-filling the system with 100 mM NaCl to remove any air bubbles. Then we introduce the particle suspension and start recording the flow rate and video footage synchronously. We allow the experiment to run until either (a) all 40 channels have clogged or (b) at least 10 hours have passed -- whichever comes first. 

Each video is then reviewed to extract the time and location of each clog. There are 40 channels with 20 constrictions per channel, or 800 potential clog locations. To aid in the review process, we utilize a grid overlay, custom-made for this device footage. This makes the review process simpler and less error-prone.

For the experiments performed in the present study, the flow is laminar with an associated Reynolds number ${\rm Re} = uL/\nu \leq 0.4$, where $u$ is the flow velocity in an open channel, $L$ is the hydraulic diameter of a pore, and $\nu=\eta/\rho$ is the kinematic viscosity of the suspension. The average Womersley number ${\rm Wo} = L\sqrt{2\,\pi\,f/\nu}$ remains smaller than 0.01 across experiments, thus transient inertial forces due to pulsation are negligible compared to viscous forces.\cite{Womersley1955} For low-Reynolds and low-Womersley number flows, such as those featured in the present study, the pressure and flow rate are linearly related through the Hagen-Poiseuille equation. \cite{OzdincCarpinlioglu2001ATopics} The average P\'eclet number is ${\rm Pe}=u\,L/D \sim 10^5 $, where $D$ is the mass diffusivity for a particle of diameter $d=2\,\mu{\rm m}$, given by the Stokes-Einstein equation $D=k_bT/(3\pi \eta d)$, where $k_b$ is the Boltzman constant and $T$ is the temperature. Therefore, in the present system, the particle transport is governed by advection, and diffusion plays a negligible role.


\section{Results}  \label{sec:Results}

\subsection{Clogging and Flow Reduction under Steady Flow}

\begin{figure*}
\centering
\includegraphics[width=1\textwidth]{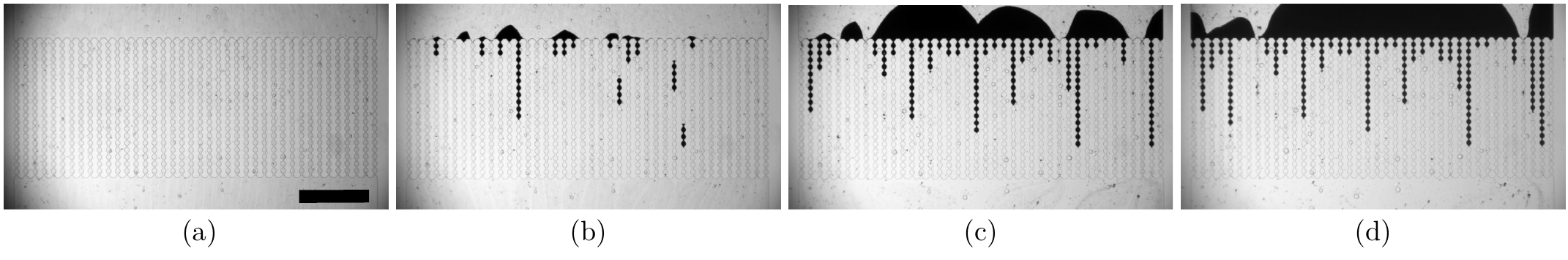}
  \caption{Time lapse showing an example of clogging dynamics observed in a steady pressure experiment, beginning with a clean device. Over time, clogs form at the constrictions while the filter cakes grow until all or most channels are clogged. Pictures taken at (a) $t=0\,{\rm s}$, (b) $1920\,{\rm s}$, (c) $11520\,{\rm s}$ s, and (d) $30000\,{\rm s}$, respectively. Scale bar is 1mm.}
  \label{fgr:Figure_2_AS}
\end{figure*}

First, we aim to characterize clogging under steady flow with $P = P_0 = 150\,{\rm mbar}$. We chose this value for $P_0$ to ensure that we do not exceed the deformation limit for the device during pulsatile experiments, which are subject to higher instantaneous pressures than the steady case (see Supplemental Materials). During the first one thousand seconds, initial clogs begin to seed the array, as shown in figure \ref{fgr:Figure_2_AS}. Since we work at constant input pressure, and the hydraulic resistances of the tubings and the reservoir are negligible compared to the microchannels, the flow rate is the device is the sum of the flow rate in each individual microchannels. In addition, the flow rate $q$ in each microchannel is drastically reduced for clogged constrictions, but remains nonzero due to the porosity of the clog.\cite{Sauret2018} As a result of the pressure-driven flow, the flow rate in the non-clogged microchannels remains similar throughout the experiments. At clogged channels, since particles may arrive but not pass, a filter cake begins to develop at each clogged constriction. This filter cake can then grow and merge with other filter cakes, or even spill into an adjacent channel, potentially catalyzing another clog. This interaction between the parallel microchannels and filter cakes is present in many real filtration environments. Eventually, most or all of the channels clog, reducing the flow rate to <5\% of the initial flow rate.

At the pore scale, we can observe the progression of an individual clog, as shown for instance in figure \ref{fgr:Figure_3_AS}. The particles successively deposit onto the walls of the constriction, shrinking its effective width and height. Eventually, a final particle or aggregate sieves the constriction, preventing future particle passage. This process takes place over hundreds to thousands of seconds for each microchannel. The clog formation depends on the concentration of aggregates in the suspension, the initial deposition of the particles on the wall, as well as the probability of sieving and bridging at the constrictions with reduced widths.\cite{sauret2014clogging,Delouche2020DynamicsAggregates}

\begin{figure}
\centering
\includegraphics[width=0.45\textwidth]{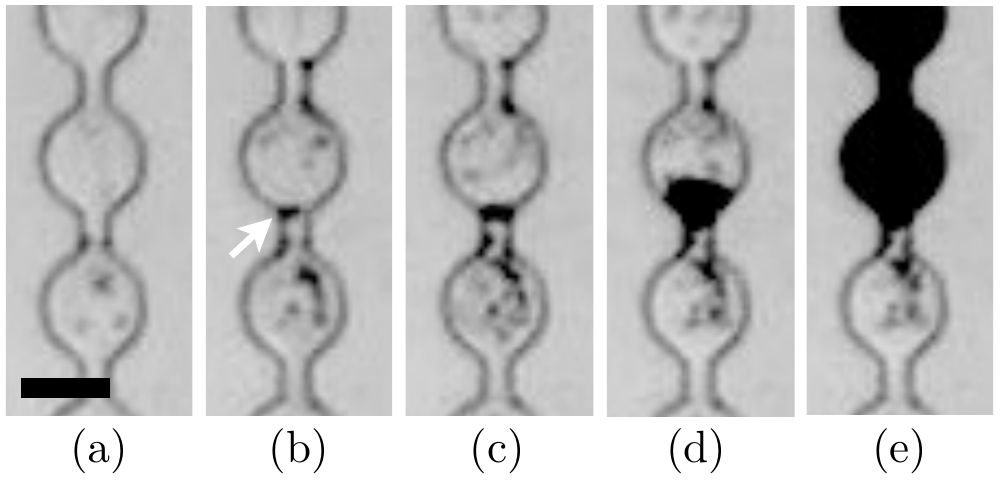}
  \caption{Time lapse showing the progression of an individual clog at the pore scale. Particles deposit along the device walls shrinking the constriction (for instance as shown by the white arrow in (b)) until a final particle or aggregate arrives and clogs the constricted channel leading to the creation of a filter cake and a reduction of the flow rate to nearly zero. Pictures taken at (a) $t=0\,{\rm s}$, (b) $300\,{\rm s}$, (c) $600\,{\rm s}$, (d) $1000\,{\rm s}$, and (e) at the end of the experiment, respectively. Scale bar is 40 um.}
  \label{fgr:Figure_3_AS}
\end{figure}

\begin{figure}[h]
\centering
\includegraphics[width=.5\textwidth]{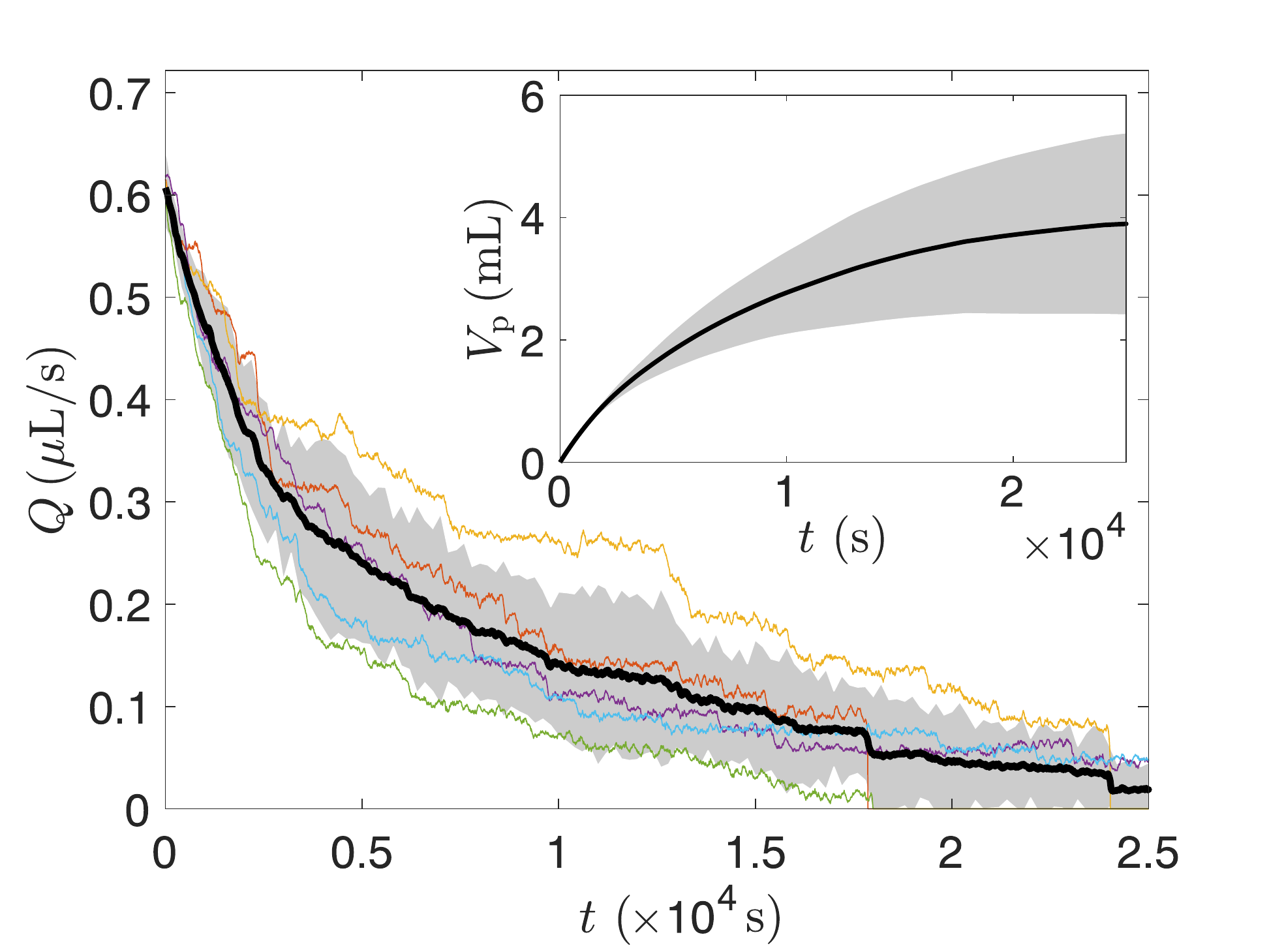}
  \caption{Flow rate $Q$ from five independent steady flow experiments (color lines). The thick black line shows the mean value, and the standard deviation among this set is represented by the gray area. Inset: Time evolution of the volume of suspension processed $V_{\rm p}=\int Q\,{\rm d}t$ obtained from integrating the mean flow rate data shown in the main panel.}
  \label{fgr:Figure_3}
\end{figure}

\begin{figure}[h]
\centering
\includegraphics[width=.5\textwidth]{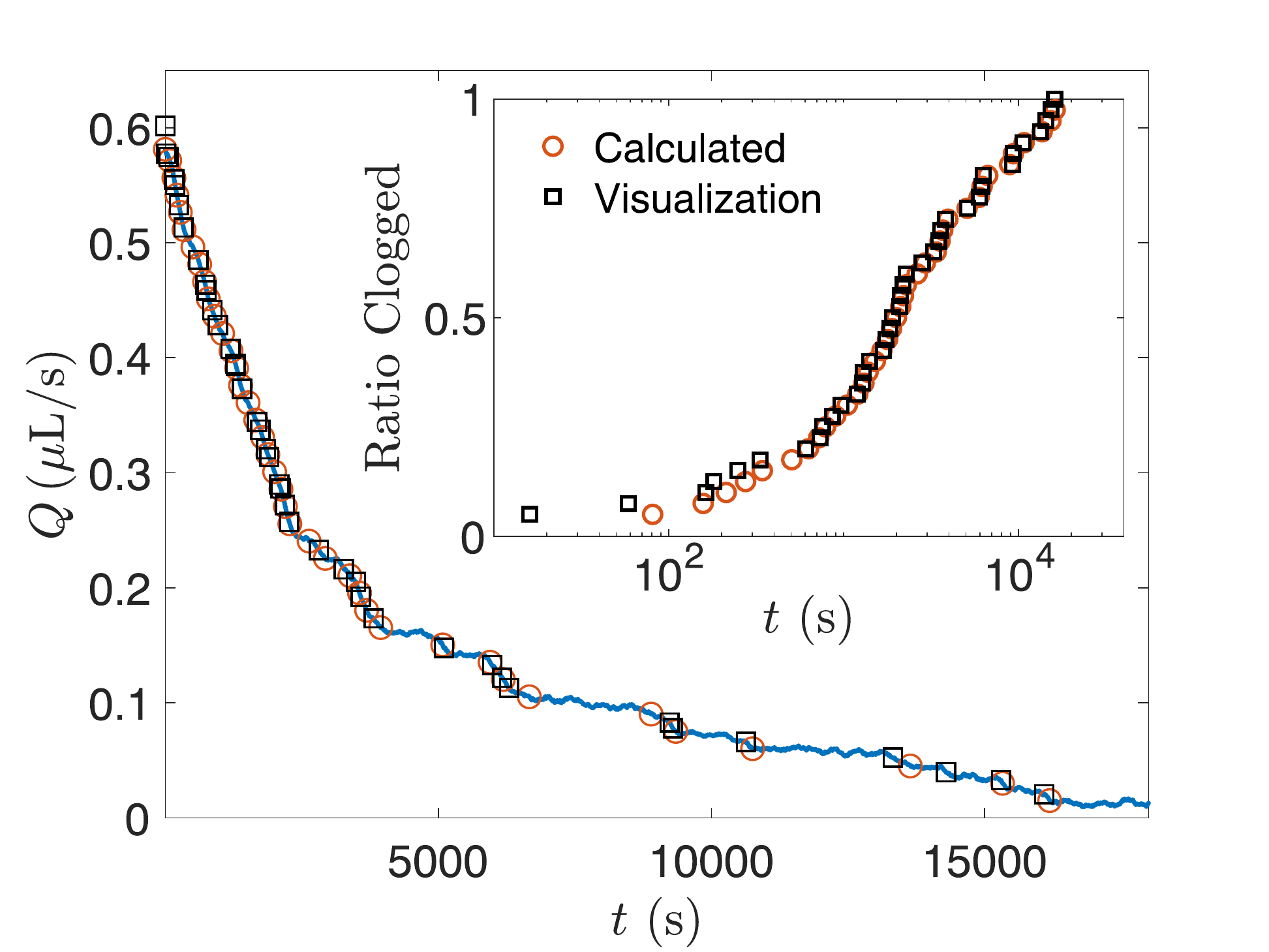}
  \caption{Comparison of raw clogging times obtained by direct visualization at the pore scale (black squares) and the calculated clogging times obtained by using a criterion on the reduction of flow rate by $1/N$, where $N=40$ is the number of parallel microchannels (orange circles). Inset: Time evolution of the ratio of clogged channel for the experiment pictured in the main panel.}
  \label{fgr:Figure_4}
\end{figure}

The flow rate for five independent steady experiments is reported in figure \ref{fgr:Figure_3}, as well as the mean value and standard deviation between this set. Due to slight variations in channel depth between microfluidic chips, the initial flow rate before any clogging occurs is equal to $Q_{\rm avg} =0.60 \pm 0.04$ $\mu{\rm L.s^{-1}}$ at 150 mbar, \textit{i.e.}, can vary slightly between devices. Therefore, for systematic comparison and to average across experiments, we report the rescaled flow rate $(Q_{\rm exp}/Q_0)\,Q_{\rm avg}$, where $Q_{\rm exp}$ and $Q_0=Q_{\rm exp}(t=0)$ are the flow rate directly measured from the flowmeter and the initial flow rate in a given device, respectively, and $Q_{\rm avg}$ is the average flow rate at time $t=0$  between all devices tested with the same pulsatile conditions. We later refer to this quantity as $Q$. We observe the same overall behavior for the five experiments reported here: the total flow rate $Q$ exhibits a fast decrease as initially, all the channels may be subject to clogging. After an initial period of frequent clogging events and a fast decrease of the flow rate $Q$, the last clogging events are slower so that over time the evolution of the flow rate slows down until reaching a value close to zero (but not zero) at the end of the experiments where most of the $N=40$ channels, if not all, have clogged. 

We define the total volume processed as $V_{\rm p}=\int\,Q\,{\rm d}t$, where we use the flow rate averaged across experiments. The aim of a method to delay clogging would be to reach a larger volume processed over the lifetime of the device. We report the evolution of $V_{\rm p}$ as a function of time in the inset of figure \ref{fgr:Figure_3}. After an initially large volume is processed, $V_{\rm p}$ saturates when most of the microchannels are clogged.

To correlate the evolution of the total flow rate through the device with the formation of individual clogs, we compare flow rate measurements with direct visualizations at the pore scale. Figure \ref{fgr:Figure_4} shows the flow rate $Q$ for a single experiment. We report the clogging times determined through direct visualization (black squares) superimposed to the evolution of the flow rate. A channel is considered fully clogged when particles can no longer be seen exiting a constriction. We can also detect clogging events based on the total flow rate. Indeed, there are $N=40$ channels per device, each with a flow rate $q=Q_0/N$. We plot in figure \ref{fgr:Figure_4} successive $1/40^{th}$ reductions in $Q_0$, assuming that it gives a good estimate of the clogging dynamics. One can note that the flow rate-predicted clogging times align fairly well with the direct visualization at the pore scale. The observed and the flow rate-predicted clogging times are re-plotted in the inset of figure \ref{fgr:Figure_4} as a ratio of channels clogged against time. The evolution with time of the relative number of clogged channels has a behavior similar to what was reported in previous studies.\cite{Wyss2006} The slight delay between an estimate based on the flow rate is likely due to the time required to build a sufficiently long filter cake, so the flow rate in the channel becomes negligible, typically a few tens of seconds.\cite{Sauret2018} The agreement between these curves demonstrates the clear relationship between flow rate and clogging times. Thus we can reasonably infer clogging times at the microchannel scale based on reductions in flow rate at the device scale.

\subsection{Influence of the Pulsatile Amplitude}

\begin{figure}[h!]
\centering
 \includegraphics[width=.5\textwidth]{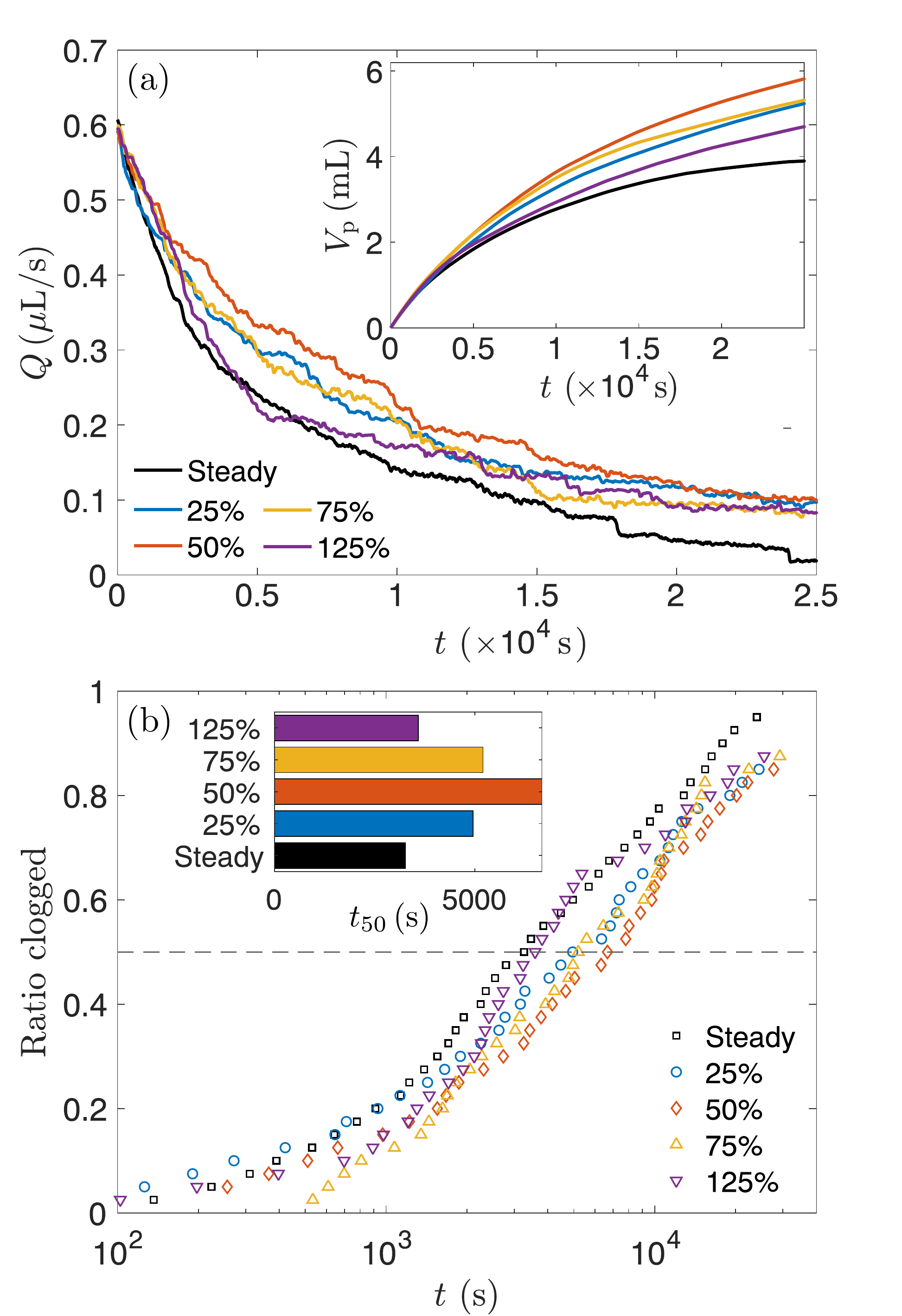} 
  \caption{Influence of the rescaled amplitude of pulsation $\delta P/P_0$ for a frequency $f=0.1\,{\rm Hz}$. (a) Flow rate for the steady experiments (in black) and for the pulsatile experiments for an amplitude of pulsation $\delta P/P_0=25\%$ (blue line), $\delta P/P_0=50\%$ (orange line), $\delta P/P_0=75\%$ (yellow line), and $\delta P/P_0=125\%$ (purple line). Inset: Cumulative volume processed $V_{\rm p}$ for pulsatile experiments, compared to steady. The color code is the same as the one used in the main panel. (b) Ratio of clogged channels for pulsatile experiments compared to the steady case. The horizontal dashed line indicates the device half-life. Inset: Average time $t_{50}$ to reach a clogging ratio of 50\% for a given amplitude of pulsation $\delta P/P_0$.}
  \label{fgr:Figure_5}
\end{figure}

For the pulsatile experiments, the maximum values for $\delta P$ and $f$ were limited by the system. At pressures above 350 mbar, the pressure-flow rate relationship becomes non-linear as the PDMS begins to deform.\cite{gervais2006flow,hardy2009deformation} At frequencies greater than 0.1 Hz, the maximum and minimum flow rates predicted by the Hagen-Poiseuille equation are not achieved due to the compliance of the system. The details of the characterization of the limits of our system are reported in Supplementary Informations. 

We vary the pulsatile amplitude at a constant frequency $f=0.1$ Hz, still keeping the average pressure constant and equal to $P_0=150\,{\rm mbar}$, and compare the results to the steady experiments. We consider pulsatile amplitudes of $\delta P = [0.25, 0.5, 0.75, 1.25]\,P_0$, corresponding to $\delta P=[37.5, 75, 112.5, 187.5]\,{\rm mbar}$. Note that the case $\delta P=1.25\,P_0$ constitutes a case where a total flow reversal occurs. The flow rate $Q$ and volume processed $V_{\rm p}$ are plotted in figure \ref{fgr:Figure_5}(a). We observe some degree of clog mitigation for every pulsatile amplitude tested; up to around 50\% improvement in flow rate and volume processed compared to steady flows. Therefore, these results suggest that simply adding a fluctuating component to the flow in an array of parallel microchannels can significantly increase the lifetime of such devices. The increase in the lifetime of the device is related to the increase in clogging time interval. Indeed, figure \ref{fgr:Figure_5}(b) reports the time evolution of the clogging ratio for each pulsatile amplitude. We also extracted the device half-life, \textit{i.e.} the expected time to reach a clogging ratio of 50\%. We found pulsation can achieve nearly 100\% improvement in filter half-life compared to a steady flow. Interestingly, even relatively small pulsation amplitude ($25\%$) increases the lifetime of the microfluidic device. However, for pulsation amplitudes leading to flow reversal, $\delta P >P_0$, the advantage of pulsatile flow on clogging mitigation decreases for the present system of parallel microchannels.

The improvement in clogging resilience thanks to pulsatile flow is likely due to several factors, some of which have been observed directly and will be discussed in section \ref{sec:discussion}. In particular, the introduction of pulsation leads to a mechanism where particles can attach during low shear and then detach at a higher shear, increasing the time it takes for an individual clog to form.\cite{Schwarze2019AttachmentFlow} PDMS is also a slightly deformable material, thus it is possible to squeeze a rigid particle or aggregate through a deformable constriction at a higher pressure,\cite{bielinski2021squeezing,zhang2017droplet,duchene2020clogging} though we note this occurrence to be extremely rare at the pressures considered here. Finally, we have observed that the filter cake itself is deformable, as it tends to shrink under higher pressures and grow under lower pressures due to variations in the local porosity. This can result in the unclogging of a channel if a filter cake can shrink enough to pass a constriction, or conversely, expand enough to facilitate a higher permeability and increased drag environment, which can move particles.

However, as emphasized previously, the largest amplitude demonstrated the least improvement in our configuration. Indeed, at amplitudes where $\delta P > P_0$, the flow is reversed with each period of oscillation. Although this condition may lead to favorable effects in a straight channel, it appears to be undesirable in parallel multichannel systems, as flow reversal allows clogs and aggregates to affect other channels. Indeed, we have observed channels becoming clogged due to a re-suspended aggregate entering adjacent microchannels. This situation will be discussed further in section \ref{sec:discussion}. In summary, pulsatile flows results in a dynamic force environment that is not present in steady flows. Generally, the unclogging phenomena out-pace the clogging phenomena for the amplitude and experimental conditions considered here since pulsatile flows lead to longer clogging times.

\subsection{Influence of the Pulsatile Frequency}

\begin{figure}[h!]
\centering
\includegraphics[width=.5\textwidth]{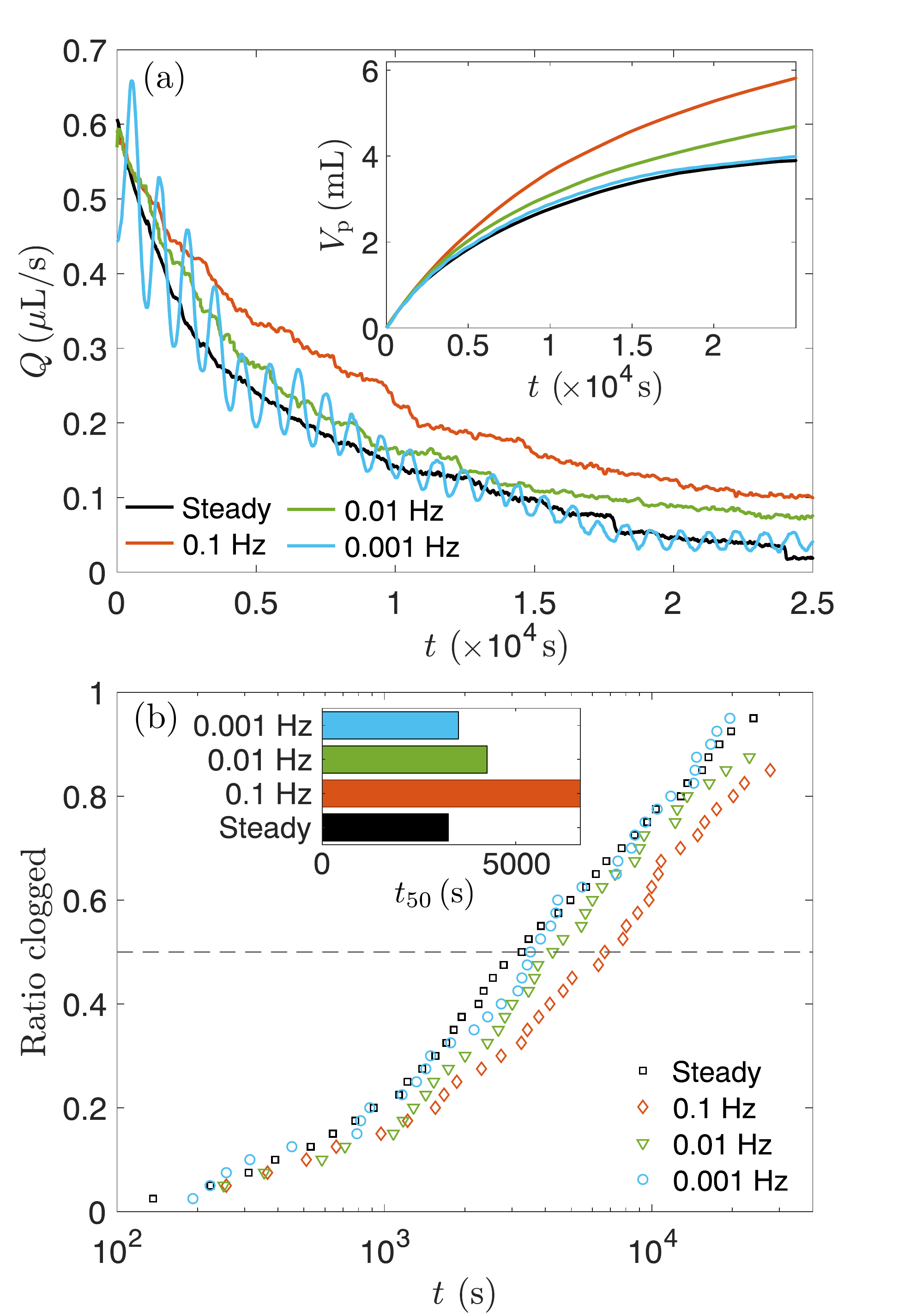}
  \caption{Influence of the frequency of pulsation $f$ for an amplitude $\delta P/P_0=50\%$. (a) Flow rate for the steady experiments (in black) and for the pulsatile experiments ($\delta P/P_0=50\%$) for a frequency $f=0.1\,{\rm Hz}$ (orange line), $f=0.01\,{\rm Hz}$ (green line), and $f=0.001\,{\rm Hz}$ (blue line). Inset: Cumulative volume processed $V_{\rm p}$ for pulsatile experiments, compared to steady. The color code is the same as the one used in the main panel. (b) Ratio of clogged channels for pulsatile flow experiments compared to the steady case. The horizontal dashed line indicates the device half-life. Inset: Average time $t_{50}$ to reach a clogging ratio of 50\% for a given frequency of pulsation $f$.}
  \label{fgr:Figure_6}
\end{figure}

A second key parameter that characterizes pulsatile flows is the frequency $f$. We now keep the pulsatile amplitude constant, $\delta P = 0.5\,P_0$ and vary the frequency $f$ between $10^{-3}\,{\rm Hz}$ and $10^{-1}\,{\rm Hz}$. The flow rate and the total volume processed are reported in figure \ref{fgr:Figure_6}(a). Similar to the previous section, we also report the clogging ratio to compare the different cases in figure \ref{fgr:Figure_6}(b). Here the transition in the evolution of the clogging dynamics from steady to pulsatile flows is clear. Indeed, we observe that clogging is best mitigated at the highest frequency that we can reach in our system ($f=0.1\,{\rm Hz}$). 

At $10^{-3}\,{\rm Hz}$, the lowest frequency tested, all curves closely follow the steady flow experiments. This is because, at this frequency, the period of pulsation $T=1/f=1000\,{\rm s}$ is too large compared to the timescale of clogging, of order a few $100$s of seconds. The benefits of pulsation are therefore not observed because the device clogs as if the flow was steady. Indeed, as soon as a clog forms, a filter cake emerges. The filter cake grows, increasing its size as particles arrive but cannot pass.\cite{Sauret2018} Increasing the length of the filter cake reduces the flow rate and significantly damps the shear oscillations through the cake and the possibility of local reorganization that would be able to unclog the system.


\section{Discussion}   \label{sec:discussion}


\subsection{Probability of clogging with pulsatile flows}

\begin{figure}
\centering
\includegraphics[width=0.475\textwidth]{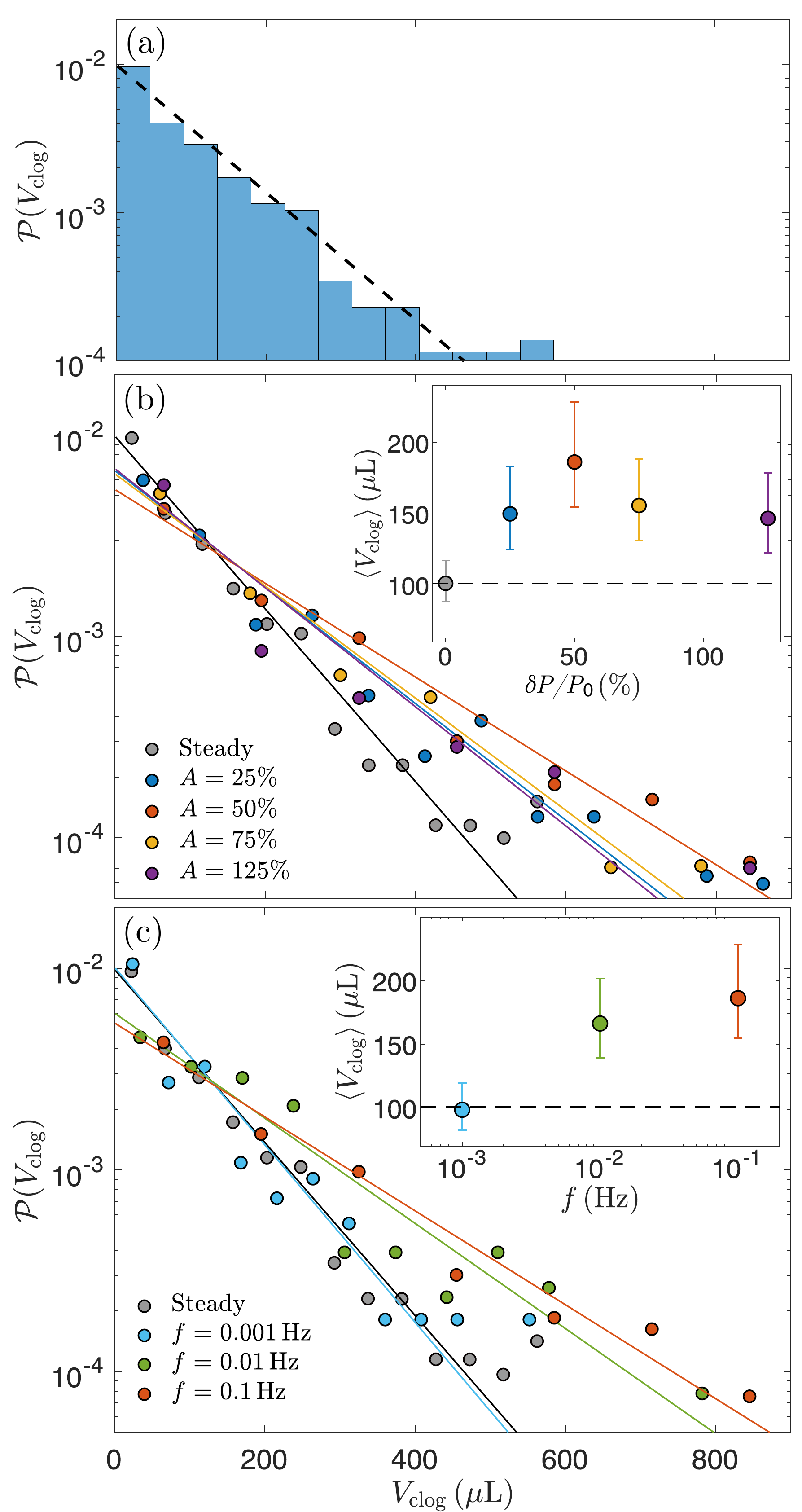}
  \caption{(a) Distribution of clogging volume intervals $\mathcal{P}(V_{\rm clog})$ in the steady case ($f=0$). The black dotted line is the best fit assuming an exponential distribution with $\langle V_{\rm clog} \rangle = 101\,\mu{\rm L}$. (b)-(c) Comparison of the distribution of clogging volume intervals $\mathcal{P}(V_{\rm clog})$ when varying (b) the amplitude of pulsation $\delta P/P_0$ at $f=0.1\,{\rm Hz}$ and (c) the frequency of pulsation $f$ at $\delta P/P_0=0.5$. In both figures the circles are the experimental results and the continuous line are the corresponding best fit with an exponential distribution given by Eq. (\ref{eq:exponential_distribution}). In figures (b) and (c) the insets show the average clogging volume interval $\langle V_{\rm clog} \rangle$ and the horizontal dashed line indicates the steady case.}
  \label{fgr:Figure_7}
\end{figure}

In this system, the clogging of a pore is due to particle deposition, aggregation, and sieving of large particles. In each experiment, an 8 $\mu$m filter sets the cutoff size for large particles and aggregates. Note that non-spherical particles, for instance, aggregates made of several particles, can have a length larger than $8\,\mu{\rm m}$ and a width smaller than $8\,\mu{\rm m}$. These aggregates or large particles could still flow through the filter and later clog the microchannels. Such aggregates are likely to be found in infinitesimal quantity but can still control the clogging events.\cite{Delouche2020DynamicsAggregates,Delouche2021TheScale} Since the constrictions are of width $\sim 10 \,\mu{\rm m}$ and the particles have a diameter $\sim 2.4 \,\mu{\rm m}$, this means that sieving is not possible until enough particles have deposited onto the sidewalls of constrictions. We observe that particle-wall attachment begins immediately after starting an experiment, forming a sparse monolayer. Particle-particle attachment is minimal due to their electrostatic stabilization. This means that after the deposition of a monolayer, the sieving of aggregates of superparticles that can be smaller than $8 \,\mu{\rm m}$ governs the clogging process since locally the width of the constriction can be reduced down to $5-6 \,\mu{\rm m}$. These aggregates, or large particles, may be anisotropic; thus, they may sieve in some orientations and flow in others.\cite{Delouche2020DynamicsAggregates} 

Sieving in an array of parallel microchannels can be described by a Poisson process. \cite{sauret2014clogging,Sauret2018} In the present case, we assume that (i) large aggregates or particles are present in the suspension, which can clog channels by sieving after monolayer deposition, (ii) the deposition of the monolayer happens on a timescale smaller than the sieving (\textit{i.e.}, for a dilute amount of large aggregates in the suspension) and that the monolayer appears at the same rate for every channel. Finally, (iii) each free constriction has an equal probability of intercepting an aggregate that may sieve it. In addition, we consider that the clogging dynamics follows a Poisson process for steady and pulsatile forcing. The pulsation can decrease the possibility of clogging through different mechanisms such as the reorientation of a non-spherical aggregate, the temporary increase of pressure on the sieved particle that eventually unclogs the channel, etc. Therefore, we define $c$ as the number of clogging entities per unit volume (due to large aggregate, bridging events, etc.). Note that $c$ is not simply a concentration of large aggregates in the suspension like in previous studies,\cite{sauret2014clogging,Delouche2020DynamicsAggregates} but also depends on the pulsation parameters.

The clogging time interval in seconds between the clogging of the $i$-th and the ($i-1$)-th channel is defined as $t_{{\rm clog},i}=t_i-t_{i-1}=1/[c\,Q(i)]$. \cite{sauret2014clogging} The index $i \in [1,\,N]$ therefore denotes the $i$-th clogging event, irrespective of the position of the channel in the microfluidic array. The clogging time interval depends on $c$, and on the instantaneous total flow rate $Q(i)=(N-i)\,q$, where $N-i$ is the number of unclogged channels and $q$ is the flow rate in each unclogged channel. Note that since the hydraulic resistance of the reservoir and the tubings are negligible compared to the microchannels, $q$ depends at first order on the pressure input $\Delta P(t)$ and the hydraulic resistance of a single microchannel $R_{\rm hyd}$ through $q=\Delta P/R_{\rm hyd}$. The average clogging time $\langle t_{{\rm clog}} \rangle $ is then given by:\cite{Sauret2018}
\begin{equation}
\left\langle t_{\operatorname{clog}}\right\rangle=\frac{1}{N} \sum_{i=0}^{N-1} \frac{R_{\rm hyd} /(N-i)}{c \,\Delta P}=\frac{1}{N q c} \sum_{j=1}^{N} \frac{1}{j}.
\end{equation}
For a number of channels $N$ large enough, this expression leads to

\begin{equation}
\left\langle t_{\mathrm{clog}}\right\rangle=\frac{H_{N}}{N\,c\,q},
\end{equation}
where $H_N = \ln N + \gamma$ is the harmonic number
of order $N$, with $\gamma \simeq 0.577$ the Euler's constant. Using the time interval is convenient for steady flows, since time is linearly related to the number of particles passing through a constriction, which controls the clogging.\cite{Wyss2006,sauret2014clogging} However, with pulsatile flows, the number of particles depends on the flow rate, which varies throughout the pulsatile waveform.  Once the period of oscillation $T=f^{-1}$ is comparable to the mean clogging interval $\langle t_{\rm clog} \rangle $ a different metric must be used. We use an integral approach to describe the clogging interval as the volume of fluid that passes through the device between each clogging event, such that  $V_{{\rm clog},i}=V_i-V_{i-1}=\int_{i-1}^i {Q(i)} {\rm d}t$. The clogging time interval $t_{{\rm clog},i}$ and the clogging volume interval $V_{{\rm clog},i}$ are related through $V_{{\rm clog},i}=t_{{\rm clog},i}\,Q(i)$. With this definition, the average clogging volume interval is given by $\langle V_{{\rm clog}} \rangle = 1/c$ and is related to $\langle t_{{\rm clog}} \rangle $ through
\begin{equation} \label{eq:tclog_Vclog}
    \langle V_{{\rm clog}} \rangle =\frac{N\,q\,\langle t_{clog} \rangle }{\ln N+\gamma}.
\end{equation}
Note that $c$ is expected to be constant for a suspension, a microchannel geometry and a steady flow. However, in this work, $c$ is the concentration per unit volume of potential clogging events such that adding pulsation can change the value of $c$. Because of the relation between $t_{{\rm clog}}$ and $V_{{\rm clog}}$, the probability of a constriction clogging within a given volume interval should also be captured by an exponential distribution:\cite{sauret2014clogging}
\begin{equation}\label{eq:exponential_distribution}
\mathcal{P}(V_{clog})=\frac{1}{\langle V_{clog} \rangle }\exp \left( -\frac{V_{clog}}{\langle V_{clog} \rangle } \right) 
\end{equation}
An example of such probability distribution is plotted in figure \ref{fgr:Figure_7}(a) for the steady case. We observe that the exponential distribution given by equation (\ref{eq:exponential_distribution}) captures well the volume interval distribution with {$\langle V_{clog} \rangle =102 \pm 14\,\mu {\rm L}$} for the steady case. Note that if instead of considering $\mathcal{P}(V_{clog})$ we report $\mathcal{P}(t_{clog})$ in the present steady case, we obtain $\langle t_{clog} \rangle =598\,{\rm s}$, close to the value calculated using equation (\ref{eq:tclog_Vclog}), $\langle t_{clog} \rangle =648\,{\rm s}$.

We perform a similar analysis for the different pulsatile flows and report the distribution of clogging volume interval when varying the amplitude of pulsation $\delta P/P_0$ in figure \ref{fgr:Figure_7}(b), and when varying the frequency $f$ in figure \ref{fgr:Figure_7}(c). The inset in both figures show the corresponding average clogging volume $\langle V_{clog} \rangle $, which is related to the probability of clogging of the system.\cite{sauret2014clogging} This quantitative analysis demonstrates further that there is a noticeable improvement in clog mitigation at the highest frequency used here (around a factor 2 at $\delta P=0.5\,P_0$) and that this improvement decreases when flow reversal occurs ($\delta P=1.25\,P_0$). Besides, higher frequencies can shift the sieving probability by increasing $\langle V_{clog} \rangle $, and that the lowest frequency of $f=0.001\,{\rm  Hz}$ has a negligible effect (inset in figure \ref{fgr:Figure_7}(c)). This result is likely due to the period of pulsation, $T_{0.001 Hz} = 1000 s$, being larger than the mean clogging time for steady flows, $\langle t_{\rm clog} \rangle_{\rm steady}  \approx 630 s$.

The improvement in volume dispensed between two clogging events, and thus a decrease of the probability of clogging, is due to the reorganization of sieved particles and aggregates, as well as some slow reorganization of the filter cakes, as their shear environment changes, preventing channels from clogging or even unclogging channels which had previously become clogged. We highlight these phenomena through direct visualization at the pore scale in the following section.

\subsection{Pulsatile flow can delay clogging}

We have seen that pulsatile flows can improve the operational life of the microfluidic array. To explain these observations, we review the videos for any clogging phenomena which are unique to pulsation. Generally, pulsations can rearrange particles, and this can result in either clogging or unclogging, depending on several factors. We observed directly at the pore-scale three mechanisms that can delay clogging in pulsatile flows, highlighted in figure \ref{fgr:Figure_8}(a)-(c). The first two can be observed in all pulsatile flows considered in this study, while the last is only observed when $\delta P \ge P_0$, \textit{i.e.}, for flow reversal. 

\begin{figure}[h!]
\centering
 \subfigure[]{\includegraphics[width=0.475\textwidth]{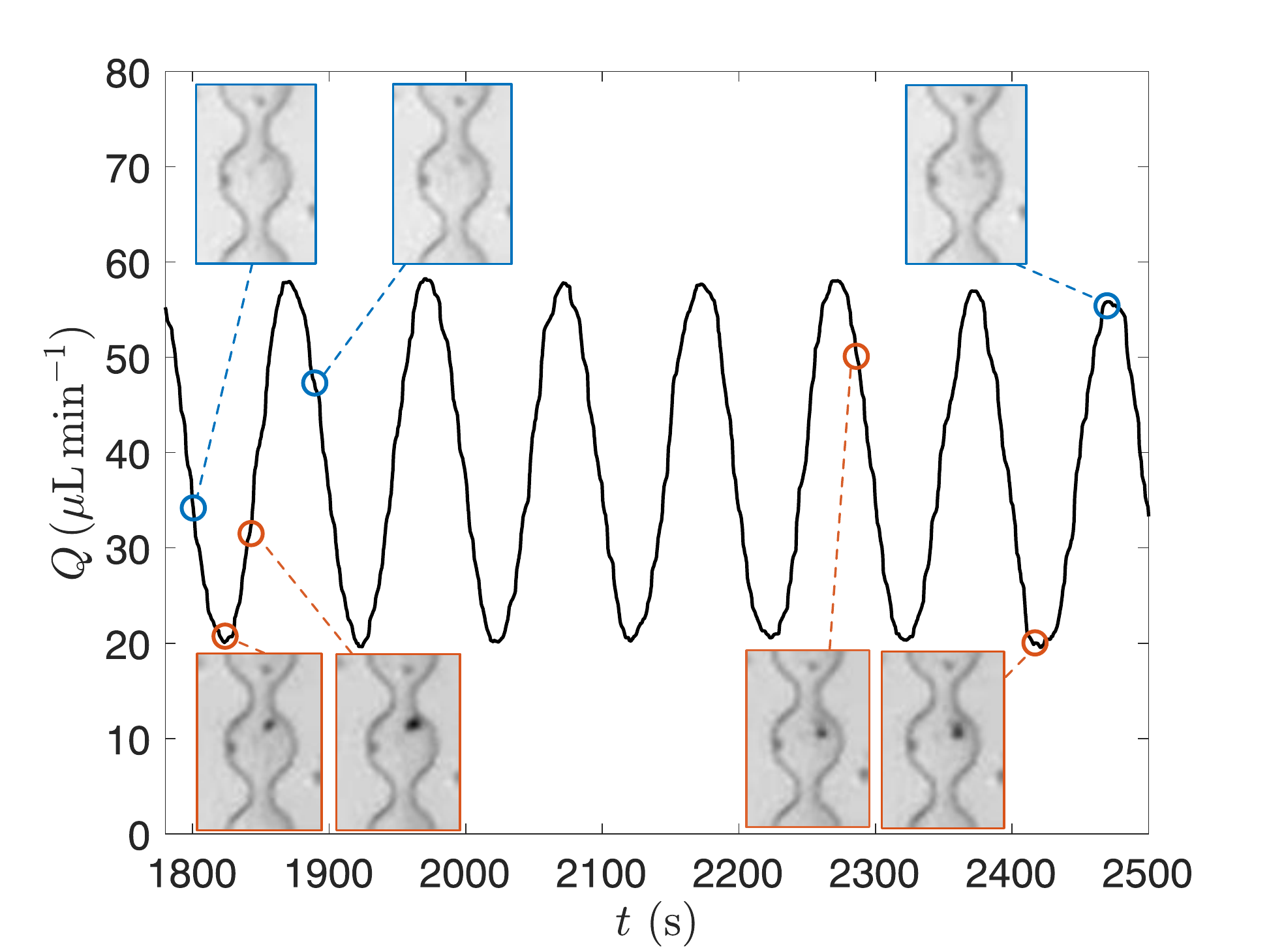}} 
   \subfigure[]{\includegraphics[width=0.475\textwidth]{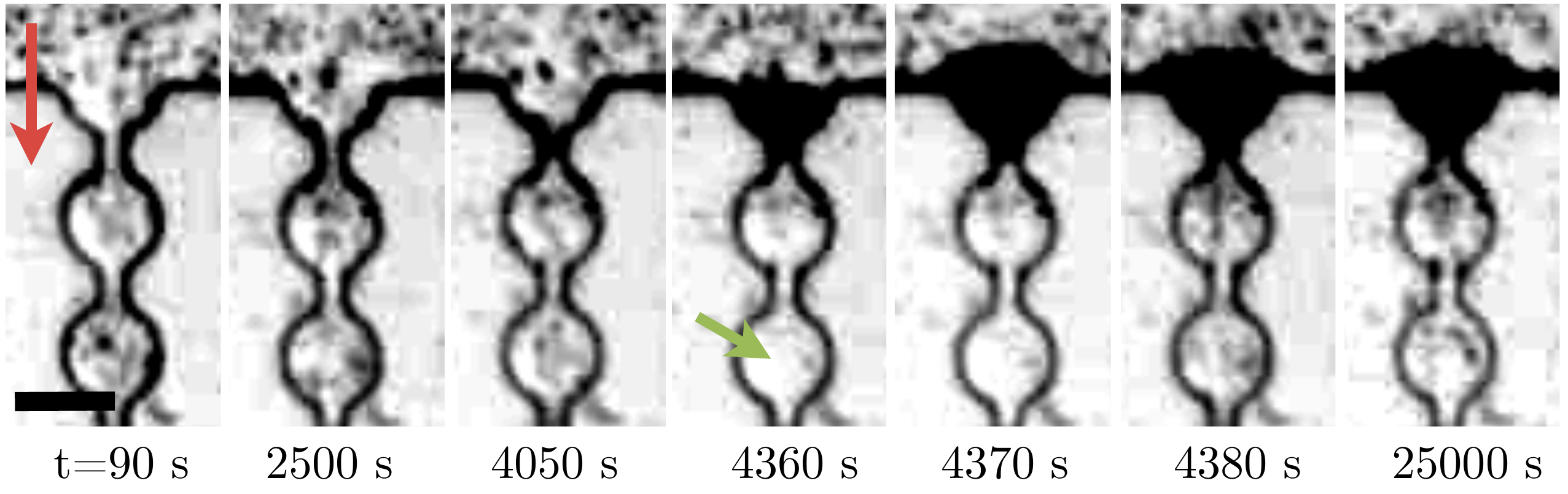}}
   \subfigure[]{\includegraphics[width=0.475\textwidth]{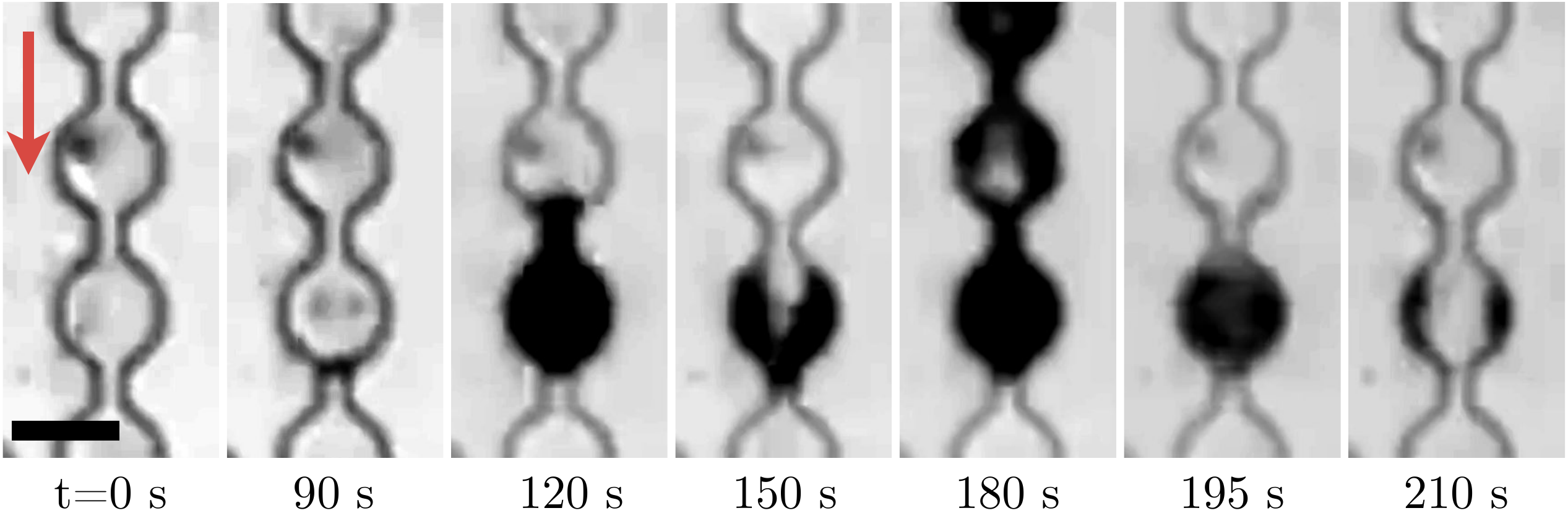}}
  \caption{Examples of unclogging via pulsatile flows. (a) Deposition and erosion of particles during pulsation. The channel begins as clear. Then particles periodically attach (red) and detach (blue) from the channel walls. (b) Unclogging of a channel without flow reversal, due to the reorganization of the filter cake at the constriction. The constriction becomes clogged at 4360 s when no particles can be seen exiting the constriction (white background shown by the green arrow). The constriction becomes unclogged at 4380 s (grey background), when particles can be seen exiting the constriction again, and remains unclogged for over 20,000 more seconds. (c) Unclogging of a channel during the flow reversal process with $\delta P=1.25\,P_0$ at $f=0.1$ Hz. The constriction clogs at $t=90 \,{\rm s}$ and the filter cake begins to grow. The flow reversal disperses and rearranges the particles in the filter cake, allowing them to reflow over several periods. The channel eventually becomes unclogged at $t=210\,{\rm s}$, with only a few particles left deposited on the sidewalls. In both figures, the scale bar is $50\,\mu{\rm m}$ and the average direction of the flow is indicated by the red arrow.}
  \label{fgr:Figure_8}
\end{figure}

First, pulsatile flow temporarily increases the flow and thus the shear rate and the ability to detach particles when subject to an elevated shear.\cite{Schwarze2019AttachmentFlow} The ability to erode deposited particles is illustrated for instance in figure \ref{fgr:Figure_8}(a), which overlays clogging images with real-time flow rate measurements over a few periods of oscillations. Particles are periodically deposited at some time and then re-suspended at a later time, maintaining a relatively clear channel when compared to steady flows. Notably, particles tend to re-suspend during high-shear conditions for each period. 

The second unclogging mechanism that we observed in our experiments involves the reorganization of the filter cake. The filter cake can compress or expand as the pressure rises and falls, respectively. \cite{Iritani2018CompressionFiltration} The pulsations also allow particles in the filter cake to rearrange, likely due to a creep phenomenon.\cite{houssais2021athermal} This rearrangement can be large enough to partially unclog a channel, as illustrated by the example in figure \ref{fgr:Figure_8}(b). The pictured channel clogs at 4360 s, indicated by the lack of particles exiting the constriction at this time. By 4380 s, particles can be seen exiting the constriction again due to rearrangement of the filter cake. In this case, particles continue to flow through the constriction even 20,000 seconds after unclogging.

The last unclogging mechanism has only been observed with flow reversal ($\delta P > P_0$), and is demonstrated in figure \ref{fgr:Figure_8}(c) for $\delta P = 1.25 P_0$ and $f=0.1\,{\rm Hz}$. A channel becomes clogged, and the filter cake begins to grow, as would be observed for any steady and pulsatile flow. However, upon flow reversal, the filter cake is ejected and then rearranged when forward flow resumes. After several periods of rearrangement, the constriction is fully unclogged, and the flow is restored. However, this scenario is observed for a minority of cases for constrictions that clog. The possible unclogging upon flow reversal is likely associated either with the reorganization and breakup of an arch of particles or with the reorientation of a non-spherical aggregate. Once the filter cake becomes large enough, the ejection and reorganization of particles become minimal during the flow reversal, and the microchannel remains permanently clogged. Higher frequency oscillations may increase the occurrence of this mechanism, as more frequent flow reversal could help prevent filter cakes from becoming too large to eject.

\subsection{Pulsatile flow can also accelerate clogging in parallel microchannels}

\begin{figure}[h!]
\centering
 \subfigure[]{\includegraphics[width=0.5\textwidth]{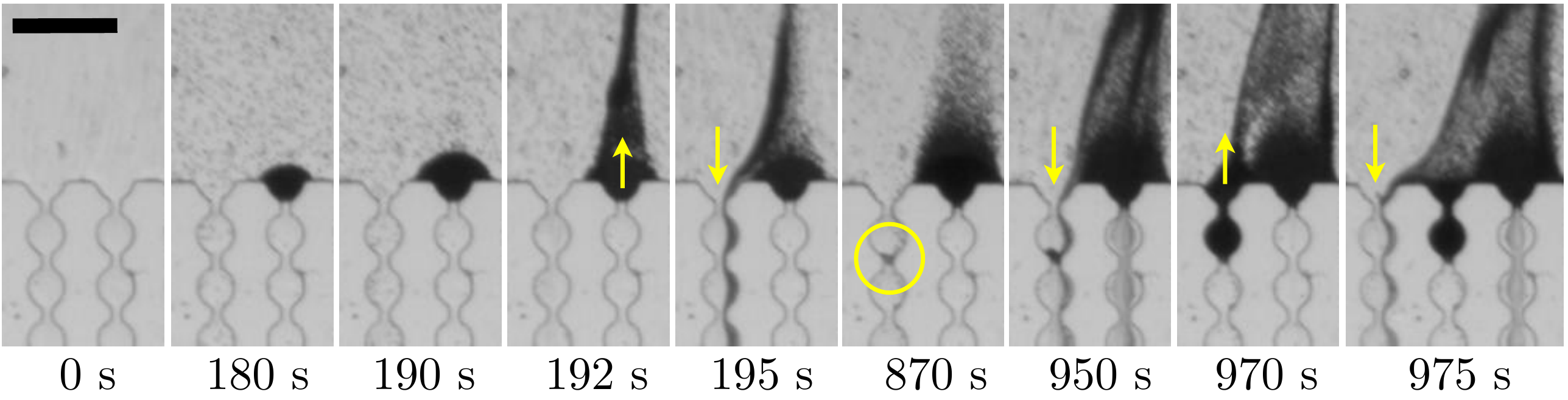}}
 \subfigure[]{\includegraphics[width=0.5\textwidth]{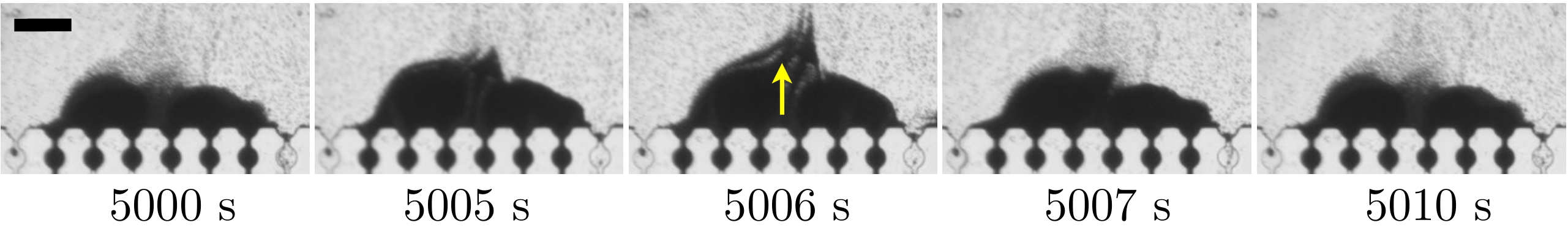}}
  \caption{Clog acceleration via flow reversal. (a) A clog forms in the right channel at $t=180\,{\rm s}$ and grows for one period of pulsation with parameters $\delta P=1.25 P_0$ and $f=0.1\,{\rm Hz}$. When the flow is reversed at $t=192\,{\rm s}$, a dense stream of particles is ejected from the right channel. Once forward flow resumes at $t=195\,{\rm s}$, the dense stream passes through the left channel (yellow arrow), locally depositing particles at a higher rate and clogging the adjacent channel (yellow circle). By $t=970\,{\rm s}$, both channels are clogged and the process continues for the next adjacent channel. (b) This effect of the flow reversal is reduced as the filter cakes become large and several adjacent channels are clogged. In one pulsation, the ejecta for this filter cake is much smaller and does not transfer to adjacent channels. Scale bar is $140\,\mu{\rm m}$ in both figures.}
  \label{fgr:Figure_9}
\end{figure}

In addition to these mechanisms that mitigate clogging with pulsatile flows, we also observe a clogging mechanism that is unique to pulsatile flows with flow reversal. In pulsatile flows without reversal, downstream particles and aggregates cannot travel upstream. With flow reversal, upstream and downstream periodically interchange. This allows filter cakes to interact with other adjacent microchannels in a unique way. A time-lapse of this behavior is featured in figure \ref{fgr:Figure_9}(a) where, during reversal, particles are ejected from the filter cake as a dense cloud ($t=192\,{\rm s}$). When forward flow resumes, many of these particles flow through a nearby open microchannel, taking the path of least hydraulic resistance (and larger local flow rate). This can accelerate the clogging, as it increases the local volume fraction of suspended particles and the number of particles flowing in an open microchannel, possibly also leading to bridging since the local volume fraction increases.\cite{Marin2018} Furthermore, if particles aggregate while compacted in the filter cake, large aggregates may be ejected and flow through an adjacent channel. If aggregates are large compared to the pore size, they cause rapid clogging through sieving.\cite{sauret2014clogging,Delouche2020DynamicsAggregates}

This peculiar dynamics explains our observations at $\delta P = 1.25 P_0$ plotted in figure \ref{fgr:Figure_5}. Due to particle resuspension during flow reversal, clogging is accelerated when compared to other amplitudes in the early part of the experiment after some filter cakes appear. However, as the experiment progresses, filter cakes grow and merge, significantly reducing the flow rate in both directions. Thus, as filter cakes grow, ejection is reduced, as illustrated in figure \ref{fgr:Figure_9}(b) where the dense cloud of particles is limited and cannot be seen entering a nearby channel. This dynamics may explain why the improvement of the flow rate at a given time upon steady after the first 5000 seconds, performing similarly to the other pulsatile experiments at 0.1 Hz since the three clog mitigation mechanisms described above continue to influence the clogging dynamics.

\section{Conclusions}

In this work, we experimentally investigate how pulsatile flows can influence the clogging rate of a dilute suspension in an array of parallel microchannels. The previous studies mentioned in Table 1 demonstrated, mainly qualitatively, that pulsatile flow is a good candidate to delay clogging but often failed to report critical pulsatile parameters or identify the relevant physical mechanisms at play. Our analysis bridges both scales while clearly reporting each pulsatile condition in a systematic investigation. By combining system flow rate measurements with direct visualization at the pore scale, we report several phenomena that mitigate or accelerate clogging and are unique to pulsatile flows. We find that all pulsatile amplitudes investigated increased the throughput over time when compared to steady flow conditions. We use the flow rate measurements to extract the clogging dynamics for each pulsatile condition and define a filter half-life as the average time required to reach a clogging ratio of 0.5, when half of the channels are clogged. In the optimal case for the present configuration, at $\delta P = 0.5 P_0$ and $f=0.1\,{\rm Hz}$, we show that pulsatile flows can nearly double the filter half-life. We attribute this to a variety of unclogging mechanisms, which we identified through pore-scale visualization. With pulsation, individual particles and aggregates may periodically detach from the device walls as their shear environment changes, delaying clogging by aggregation. Additionally, the slow rearrangement of particles in a filter cake during pulsation can result in a partial flow restoration. These phenomena are not present in steady flows.

When we consider flow reversal, where $\delta P = 1.25 P_0$, we report a unique unclogging mechanism, where sufficient flow reversal can re-suspend particles in a filter cake, unclogging the constriction. Despite this, we observe little improvement in filter half-life when compared to steady flows. This is due to the fact that adjacent channels can influence each other during flow reversal.

While the pulsatile amplitude determines which clogging and unclogging mechanisms are present, the pulsatile frequency determines the probability of these events. In our case, the highest frequency investigated ($f = 0.1\,{\rm Hz}$) yields the greatest increase in filter half-life. At the lowest frequency we investigated ($f = 10^{-3}\,{\rm Hz}$), performance is indistinguishable from the steady case. Therefore, it is critical that the frequency of pulsation is compared to the average clogging rate for a given system. If the frequency is too low, pulsation cannot delay clogging. While higher frequencies may achieve further improvement, we are unable to investigate this situation because of the limitation in our system.

This study demonstrates that even relatively weak pulsations may significantly delay clogging. In the specific case of filtration where constrictions are closely arranged in an adjacent configuration, it may be advisable to avoid flow reversal, as we observe minimal improvement in filter half-life despite the increased energy cost of large pulsations. However, in the single-channel case, flow reversal may prove highly beneficial due to its unique ability to erode deposited particle.

\section*{Author Contributions}
B.D, E.D. and A.S. designed the experiments; B.D. performed
experiments; B.D., C.T., E.D., A.S. analyzed the data; E.D. and A.S. supervised the overall project. All authors discussed the results. B.D., E.D., A.S. wrote the initial version of the manuscript. All authors proof checked the manuscript.

\section*{Conflicts of interest}
There are no conflicts to declare.

\section*{Acknowledgements}
The authors acknowledge the use of the Microfluidics Laboratory within the California NanoSystems Institute, supported by the University of California, Santa Barbara and the University of California, Office of the President. A portion of this work was performed in the UCSB Nanofabrication Facility, an open access laboratory. ED acknowledges the support from the UCSB Academic Senate Faculty Grant, and AS the partial support from the U.S.-Israel Binational Agricultural Research and Development Fund (BARD) US-5336-21.


\balance

\bibliography{biblio} 
\bibliographystyle{unsrt}

\end{document}